\documentclass[10pt, a4paper]{article}

\usepackage{lrec2026} 

\usepackage[most]{tcolorbox}

\usepackage{textcomp}
\usepackage{booktabs}  
\usepackage{array}      
\usepackage{tabularx}   
\usepackage{ragged2e}  
\usepackage{makecell}   
\usepackage{subcaption} 
\usepackage{hyphenat}

\title{Test-Time Strategies for More Efficient and Accurate Agentic RAG}

\name{Brian Zhang$^1$, Deepti Guntur$^1$, Zhiyang Zuo$^1$, Abhinav Sharma$^1$, Shreyas Chaudhari$^1$, \\ {\bf \large Wenlong Zhao$^1$, Franck Dernoncourt$^2$, Puneet Mathur$^2$, Ryan Rossi$^2$, Nedim Lipka$^2$}} 

\address{$^1$University of Massachusetts Amherst, $^2$Adobe Research \\
         Amherst, MA, USA; San Jose, CA, USA \\
         \{bszhang, dguntur, zzuo, abhinavs, schaudhari, wenlongzhao\}@umass.edu \\
         \{dernonco, puneetm, ryrossi, lipka\}@adobe.com\\}
         
\abstract{
Retrieval-Augmented Generation (RAG) systems face challenges with complex, multihop questions, and agentic frameworks such as Search-R1 \cite{jin2025searchr1trainingllmsreason}, which operates iteratively, have been proposed to address these complexities. However, such approaches can introduce inefficiencies, including repetitive retrieval of previously processed information and challenges in contextualizing retrieved results effectively within the current generation prompt. Such issues can lead to unnecessary retrieval turns, suboptimal reasoning, inaccurate answers, and increased token consumption.
\\
In this paper, we investigate test-time modifications to the Search-R1 pipeline to mitigate these identified shortcomings. Specifically, we explore the integration of two components and their combination: a contextualization module to better integrate relevant information from retrieved documents into reasoning, and a de-duplication module that replaces previously retrieved documents with the next most relevant ones. We evaluate our approaches using the HotpotQA \cite{yang2018hotpotqadatasetdiverseexplainable} and the Natural Questions \cite{naturalquestions2019} datasets, reporting the exact match (EM) score, an LLM-as-a-Judge assessment of answer correctness, and the average number of turns. Our best-performing variant, utilizing GPT-4.1-mini for contextualization, achieves a $5.6\%$ increase in EM score and reduces the number of turns by $10.5\%$ compared to the Search-R1 baseline, demonstrating improved answer accuracy and retrieval efficiency.
 \\ \newline  }

\begin{document}
\maketitleabstract

\section{Introduction}

\begin{figure}[t]
    \centering
    \includegraphics[width=\linewidth]{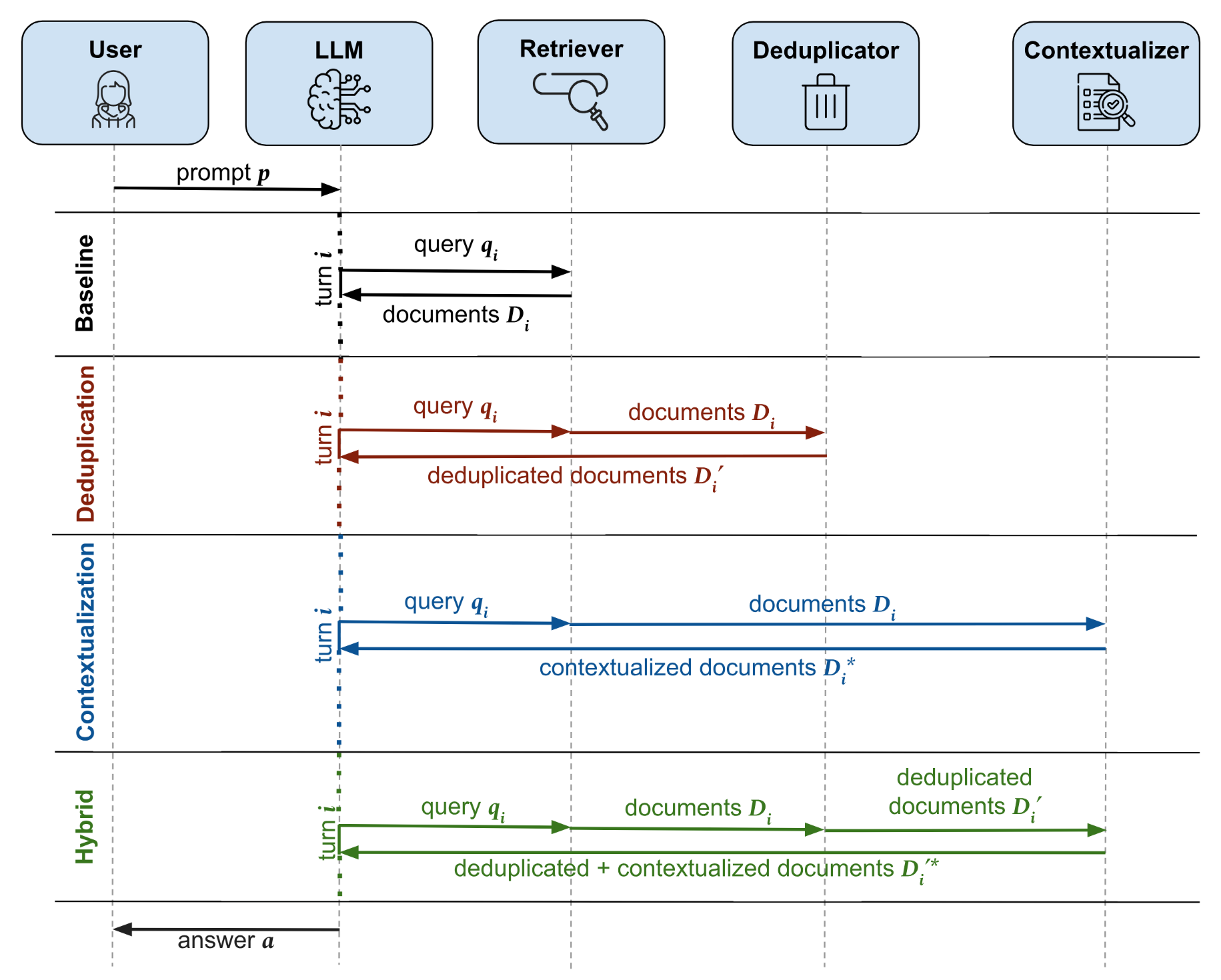}
    \caption{An illustration of the information flow for our proposed test-time strategies compared to the baseline during a single inference turn $i$. \textit{Baseline:} This represents the standard Search-R1 framework, where the LLM sends a query ($q_i$) to the retriever and directly receives the retrieved documents ($D_i$) to continue its reasoning. \textit{Deduplication:} This approach filters out previously seen content and returns only a set of novel documents ($D_i'$) to the LLM. \textit{Contextualization:} This approach parses the retrieved documents ($D_i$) and reformulates their content to improve integration into the LLM's reasoning process, returning an enhanced set of information ($D_i^*$). \textit{Hybrid:} This approach combines both modules sequentially.}
    \label{fig:approaches}
\end{figure}

RAG systems have shown promising results in complex question answering (QA) tasks by combining external document retrieval with generative language models \cite{lewis2021retrievalaugmentedgenerationknowledgeintensivenlp}. Despite this success, traditional RAG systems that rely on a single-step retrieval and generation process often struggle to handle complex or nuanced questions, especially those requiring deep contextual understanding and multi-hop retrieval. To address these complexities, recent research has proposed agentic RAG systems which utilize large language model (LLM) agents to orchestrate retrieval, refine search queries, and optimize responses \cite{singh2025agenticretrievalaugmentedgenerationsurvey, an2024goldenretrieverhighfidelityagenticretrieval, chan2024rqraglearningrefinequeries}. Another popular approach is to augment the reasoning loop of LLMs with a retrieval tool, enabling the model to autonomously use retrieval while performing multi-step reasoning \cite{li2025searcho1agenticsearchenhancedlarge, jin2025searchr1trainingllmsreason}.

A notable example of this approach is the Search-R1 framework, which uses reinforcement learning (RL) to train LLMs for interleaved reasoning and retrieval \cite{jin2025searchr1trainingllmsreason}. At inference time, the Search-R1 model first performs reasoning on a given user prompt~$p$ to either produce an answer~$a$ or generate a search query~$q$ to retrieve supporting information~$D$. More specifically, in the $i$-th turn, the query~$q_i$ is sent to a dense retriever E5 \cite{wang2024textembeddingsweaklysupervisedcontrastive}, which returns relevant passages~$D_i$ from a 2018 Wikipedia dump. $D_i$ is directly incorporated into the reasoning trace and fed back into the LLM, which continues its reasoning and repeats these steps until the final answer $a$ is produced, see Figure~\ref{fig:approaches}~(Baseline). The Search-R1 models are trained using RL methods, such as PPO and GRPO, optimizing the exact match (EM) score between the ground truth and the predicted answer.

While Search-R1 has achieved substantial improvement---up to $41\%$ over its baseline---our analysis of the Qwen2.5-7b Search-R1 model during inference has revealed several shortcomings. First, the model often performs repetitive retrieval of previously processed information, which leads to unnecessary retrieval turns, and increased token consumption and latency. Secondly, the model often struggles to effectively contextualize retrieved passages, leading to suboptimal reasoning and inaccurate answers.

\subsubsection*{Research questions} 
\begin{itemize}
\item Will a concise representation of relevant information help an LLM become more efficient and accurate in question-answering tasks?
\item Can preventing redundant document retrieval encourage greater contextual diversity, thereby improving efficiency and answer accuracy?
\end{itemize}

\subsubsection*{Proposed Approach}
Our work builds upon and extends the Search-R1 framework~\cite{jin2025searchr1trainingllmsreason} and investigates test-time approaches to improve the framework's reasoning efficiency and final answer accuracy. We address the limitations of Search-R1 through three test-time modifications that process the retrieved results~$D$: (1) a \textbf{contextualization} module, (2) a \textbf{de-duplication} module, and (3) a \textbf{hybrid} approach that combines both.

\section{Related Work}

A Memory Knowledge Reservoir \cite{shi2024enhancingretrievalmanagingretrieval} stores previously retrieved content as a title-document pair. This system first consults previously retrieved information before formulating a new query. This allows the system to produce more targeted queries. As a result, they can reduce the response time by 46$\%$ while preserving the accuracy of the response of their baseline. Rather than storing the entire document, we propose a contextualization module that prompt an external LLM to extract useful information, and incorporate this into the model's reasoning chain after each retrieval step and before the next query is generated.

Search-o1 \cite{li2025searcho1agenticsearchenhancedlarge} addresses the issue of hallucination in reasoning models by incorporating an agentic RAG mechanism to extract information from documents before integrating it into a reasoning chain.  We use a similar approach to Search-o1 by utilizing an external LLM to contextualize helpful information in a document. Unlike Search-o1 which only provides the extracted information, our pipeline retains previously contextualized information, which is passed along with newly retrieved documents at each reasoning step.

RAG-RL \cite{huang2025ragrladvancingretrievalaugmentedgeneration} introduces a reasoning language model specifically trained for RAG tasks using reinforcement learning and curriculum learning strategies. The authors demonstrate that stronger answer generation models can identify relevant contexts within larger sets of retrieved information, thus alleviating the burden on retrievers and enhancing overall performance. By benchmarking on HotpotQA and MuSiQue datasets, RAG-RL achieves performance that surpasses previous generative reader models. RAG-RL's rewards are all rule-based and determined by the final answer, output format, and the citations included. Our work only explores test-time approaches and does not involve any modifications to the model architecture or training process.

\section{Approach}

To understand the limitations of Search-R1, we conducted a qualitative analysis of Search-R1's reasoning chains. These chains contain the original user prompt, multiple turns of the model reasoning, search queries, and retrieved documents, as well as the final answer. We observed two primary limitations in the Search-R1 model. First, \textbf{Information Forgetting:} the model struggles to retain and utilize information from previous retrieval steps, often resulting in redundant or duplicate retrieval queries before arriving at a final answer. Second, \textbf{Ineffective Information Extraction: } the model often fails to effectively identify and extract the most relevant information from the retrieved documents, which hinders its reasoning and overall accuracy of the answers.

Based on these findings, we propose three test-time modifications to the Search-R1 pipeline aimed at addressing the challenges of information forgetting and ineffective information extraction. For each approach, we evaluate performance using Exact Match, LLM Match score, and the average number of retrieval steps.

\subsection{Contextualization} 
\label{sec:cont}
To assess the importance of extracting and retaining relevant information across retrieval steps, we introduce the Contextualization module, shown in Figure~\ref{fig:approaches} (Contextualization). This is an additional component in the pipeline that leverages an external language model to extract relevant information from retrieved documents and maintain a persistent memory cache based on previously contextualized information. After each retrieval step, the external LLM identifies concise, useful content and updates the cache accordingly. At each reasoning step, the model accesses both the most recently retrieved document and the accumulated cache, allowing it to reason over both new and previously retained information.

We provide the LLM with a structured prompt that instructs it to extract only the information relevant to answering the user prompt~$p$ from the newly retrieved documents~$D_i$ during each retrieval turn~$i$. This extracted content~$D_i^*$ is then appended to a persistent memory cache that accumulates across retrieval steps. The external language model is constrained to preserve all previously stored information and may only add new, relevant content. If no new helpful information is identified, the model returns the existing cache; if no cache is available, it explicitly indicates that no useful content was found. The inputs to this process include the user prompt~$p$, the newly retrieved documents~$D_i$, and the accumulated memory cache.

The use of a information cache mitigates information forgetting by retaining relevant content across retrieval steps, enabling more coherent multi-hop reasoning. Meanwhile, explicitly extracting key information from retrieved documents addresses ineffective information selection, helping the model focus on information that is most useful for answering the question.

This approach allows us to assess whether providing a concise representation of retrieved information, combined with a cache of previously relevant context can improve answer quality and reduce redundant retrievals, without modifying the underlying model. By decoupling information extraction from reasoning, it introduces an agentic component that enables more structured, context-aware inference and better utilization of retrieved knowledge across steps.

\subsection{De-duplication of retrieved documents}
\label{sec:dedup}

To investigate the causes of duplicate search query generation and evaluate the effect of retrieval redundancy on model performance, we introduce a de-duplication module that filters out documents retrieved in previous steps. This approach tests the hypothesis that the model generates repeated queries because it deems the initially retrieved information insufficient for the task. By preventing repeated access to the same content, this module encourages the model to incorporate a broader set of documents throughout its reasoning process. Specifically, when a retrieved document is discarded as a duplicate, the system is forced to consider the next-highest-ranked passage from the retriever's full ranked list. This effectively allows the model to continue to explore parts of the document collection that did not appear in the top-$k$ results of previous turns, thereby increasing the diversity of the information it considers. 

At the retrieval step of each turn, $k=3$ documents~$D_i$ are returned. However, these documents might have been already seen. In this approach, Figure~\ref{fig:approaches} (Deduplication), we maintain a set of unique document IDs for all passages seen during the reasoning processes in previous turns for a given user prompt~$p$. Any new retrieval that returns a document whose ID is already in this set is discarded and replaced by the next-highest-ranking, unseen document. The result is a set of unseen documents~$D_i'$.

Ultimately, this allows us to examine whether reducing retrieval overlap leads to improved answer accuracy and fewer redundant search queries. If information forgetting is the cause of repeated retrievals, we expect the De-duplication pipeline to result in a drop in answer accuracy, as it means the LLM no longer has access to information to answer the question correctly.

\subsection{Hybrid}
\label{sec:hybrid}
The hybrid approach combines the Contextualization with the De-duplication approaches to evaluate whether retaining extracted relevant information while enforcing retrieval diversity can jointly enhance reasoning performance. By integrating the contextualization module  with non-redundant retrieval, this setup allows us to test whether the limitations of one component (e.g., information forgetting or redundancy) can be mitigated by the other, leading to improved answer accuracy and more efficient use of retrieved content.

\section{Experiments}

\subsection{Setup}

\subsubsection{Data source}

Search-R1 \cite{jin2025searchr1trainingllmsreason} reports performance on the HotpotQA \cite{yang2018hotpotqadatasetdiverseexplainable} and Natural Questions (NQ) \cite{wang2024textembeddingsweaklysupervisedcontrastive} dataset. Since labeled test sets for these two datasets are not publicly available, we follow prior work and use the validation sets. Retrieval is performed on the 2018 Wikipedia dump with the E5 retriever.

\subsubsection{Data splits}

To reduce the cost associated with querying external LLMs, we created a smaller subset of question-answer pairs for evaluation. Specifically, we randomly sample 500 question-answer pairs from the HotpotQA \cite{yang2018hotpotqadatasetdiverseexplainable} and NQ \cite{wang2024textembeddingsweaklysupervisedcontrastive} validation sets. This subset is solely used for evaluation; no hyperparameter tuning or training is performed using this subset. All reported metrics are based on this validation set.

\subsubsection{Baselines}

We utilize the already trained Qwen2.5-7b Search-R1-base (PPO) as our main baseline for comparison. While running inference with both, Qwen2.5-3b Search-R1-base (PPO) and Qwen2.5-3b Search-R1-instruct (GRPO), the latter model exhibit difficulties in adhering to the structured output format specified by the Search-R1 framework. Inference outputs show frequent failures to generate required output tags such as \textit{<think>} and \textit{<search>} within the iterative reasoning loop. Additionally, the model often generates retrieved information under \textit{<information>} tags by itself after the search query, and also occasionally fails to produce a final answer at the end of the reasoning chain. These behaviors indicate limitations in the ability to reliably follow instruction-guided formatting. Therefore, we only enhance the superior Qwen2.5-7b Search-R1-base (PPO) model to test our modules and report the corresponding results in Table \ref{tab:em_scores}. For all approaches, we run inference on our validation dataset of 500 questions and compute exact match, LLM match, and the average number of turns.  

\subsubsection{Implementation details}

We built on top of the publicly available Search-R1 source code on GitHub, where we make modifications to the model prompt to optimize the model's behavior. For inference on the trained model, we use the HuggingFace transformer library to perform forward pass, then run the Wikipedia article chunk E5 dense retriever provided in the Search-R1 GitHub repository. Contextualization and LLM-As-A-Judge to perform LLM match is performed by GPT-4.1-mini by calling the OpenAI API with an API key.

\subsubsection{Evaluation Metrics}

The overall performance of our model is reported as the exact match (EM), just as in Search-R1. An analysis of the Search-R1 baseline reveals several false negatives where the predicted answer string does not exactly match the golden answer, despite referring to the same underlying entity, a discrepancy that is easily recognizable to human evaluators (Examples, "2" and "Two", "950 Pesos" and "P950"). In order to scale up these judgments, we prompt an external LLM model (OpenAI GPT-4.1-mini) to evaluate whether the predicted answer matches the golden answer. We call this evaluation metric LLM Match, and we present it in addition to the Exact Match metric in the results section.

For LLM Match, the model is given both the predicted answer and a set of ground truth answers, and is instructed to determine whether the predicted answer is semantically equivalent to any of the gold answers. Minor differences in phrasing are permitted as long as the predicted answer conveys the same meaning.

The prompt directs the model to assign a binary score:
\begin{itemize}
    \item \textbf{1} if the predicted answer is semantically equivalent to the ground truth, and
    \item \textbf{0} if it is incomplete or diverges in meaning.
\end{itemize}

The evaluation explicitly focuses on semantic similarity, independent of factual correctness. This setup enables scalable, consistent semantic evaluation without human annotators. In addition, since we are focused on retrieval efficiency, we report the average number of retrievals. It is important to consider this metric in the context of the EM score as the model can drive the number of retrieval iterations to 0 by always hallucinating an answer and performing no retrieval.

\subsection{Results}

\begin{figure}[ht]
    \centering
    \includegraphics[width=1\linewidth]{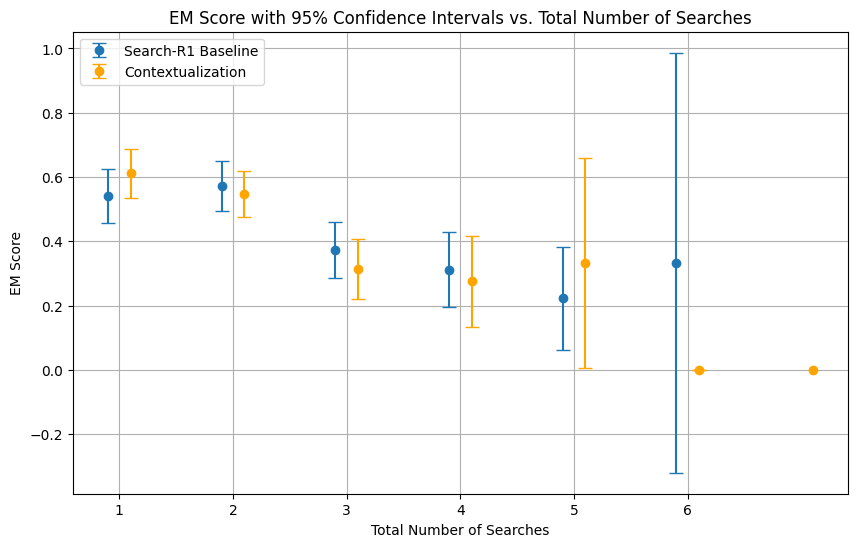}
    \caption{Illustration that questions requiring more agentic turns are inherently more difficult, as shown by the downward trend in Exact Match (EM) score for both the Search-R1 baseline and our Contextualization module. While the Contextualization module achieves a slightly higher mean EM at some points, the overlapping 95\% confidence intervals indicate that we are not seeing a statistically significant improvement or difference between the two compared approaches for any given search count.}
    \label{fig:acc-numret}
\end{figure}

\begin{table}[hbt]
    \centering
    \caption{Performance comparison of our proposed modules against the Search-R1 baselines on our 500-question evaluation set. We report Exact Match, LLM Match, and the average number of turns. Our modules are applied as test-time enhancements to the Qwen2.5-7b base (PPO) model. Cont: Contextualization (Section~\ref{sec:cont}). De-Dup: De-duplication (Section~\ref{sec:dedup}). Hybrid (Section~\ref{sec:hybrid}). Best results for our approaches are in bold.}
    \label{tab:em_scores}
    \begin{tabularx}{\linewidth}{@{} >{\RaggedRight\arraybackslash}X c c c @{}}
        \toprule
        \textbf{Variant} &
        \thead{Exact\\Match} &
        \thead{LLM\\Match} &
        \thead{avg. \#\\searches} \\
        \midrule
        \multicolumn{4}{@{}l@{}}{\textbf{Qwen2.5-3b Search-R1}} \\ \addlinespace
        base (PPO) & 0.292 & 0.356 & 1.410 \\
        instruct (GRPO) & 0.310 & 0.396 & 2.054 \\
        \midrule
        \multicolumn{4}{@{}l@{}}{\textbf{Qwen2.5-7b Search R1}} \\ \addlinespace
        base (PPO) & 0.464 & 0.538 & 2.392 \\
        base (PPO) w/ Cont \textbf{(Ours)} & \textbf{0.490} & \textbf{0.574} & \textbf{2.142} \\
        base (PPO) w/ De-Dup \textbf{(Ours)} & 0.478 & 0.560 & 2.498 \\
        base (PPO) w/ Hybrid \textbf{(Ours)} & 0.480 & 0.568 & 2.154 \\
        \bottomrule
    \end{tabularx}
\end{table}

In terms of answer accuracy, the Contextualization approach achieves a $5.6\%$ increase in EM and a $6.7\%$ increase in LLM match score compared with the Search-R1 baseline. In addition, it is also the most efficient, reducing the average number of searches to 2.142, compared with the baseline which has 2.392 searches. While the De-Duplication and Hybrid approach have around similar gains in EM and LLM match over the baseline, only the Hybrid approach has a decrease in the average number of retrievals, similar to the Contextualization. In fact, the De-Duplication pipeline is actually less efficient than the baseline, with 2.498 average retrievals compared with the baseline's 2.392 average retrievals. Overall, the Contextualization approach still achieves the highest EM, LLM match, and lowest average number of retrievals. All metrics along with the baseline are shown in Table \ref{tab:em_scores}.

We examine the outputs of the baseline Search-R1 and the De-Duplication approach to determine the source of the decrease in efficiency. We observed that the Search-R1 baseline is more likely to stop searching the same objective when its search result in duplicated documents, as they offer no new information. In contrast, the De-Duplication approach only returns new documents, causing the model to continue generating similar search queries in an effort to gather more context for the current search objective. This behavior leads to an increased average number of queries in the De-Duplication approach. However, the additional context is rarely helpful, as the necessary information is often already present in the initial retrieval but fails to be extracted by the model, resulting in only a small improvement in answer accuracy.

Figure \ref{fig:acc-numret} shows a 95$\%$ confidence interval around exact match score of the Search-R1 baseline and the Contextualization pipeline, conditioned on the total number of searches performed. While the difference between the two never appears statistically significant, we observe a downward trend in both. This would suggest that the exact match is negatively correlated with the number of retrievals. 

For the Search-R1 baselines and the three non-training approaches, the LLM match is 16 to 18$\%$ greater than the exact match score. Investigating the outputs where the LLM determines the golden and predicted answers match but exact match fails, we observe two common patterns: numerical answers and shortened or abbreviations of names.

\section{Conclusion}

In this work, we implemented and evaluated two inference-time enhancements to the Search-R1 pipeline: (1) a Contextualization module, (2) a De-duplication module for retrieved documents, and (3) a Hybrid approach combining both. All of these approaches improve the answer accuracy of the Search-R1 framework that serves as our baseline. In addition, our contextualization module also reduces the number of turns, while the De-Duplication module increases it. We evaluated a hybrid approach combining both methods, which achieved gains in both accuracy and retrieval efficiency---although not as strongly as the Contextualization module alone.



\section{Bibliographical References}
\bibliographystyle{lrec2026-natbib}
\bibliography{literature}


\end{document}